# GLAST, GRBs, and Quantum Gravity

**J.P. Norris[1], J.T. Bonnell[1], G.F. Marani[1], and J.D. Scargle[2]**
[1] *Laboratory for High Energy Astrophysics, NASA/GSFC, Greenbelt, MD 20771, USA*
[2] *Space Science Division, NASA/ARC, Moffett Field, CA 94035-1000, USA*

**Abstract**
The fast temporal structures and cosmological distances of gamma-ray bursts (GRBs) afford a natural laboratory for testing theories of frequency-dependent propagation of high-energy photons, as predicted for quantum gravity (QG). We calibrate the sensitivity of the proposed Gamma-ray Large Area Space Telescope (GLAST) by performing simulations which include: the response of GLAST to a GRB fluence distribution; a distribution of spectral power-law indices similar to the EGRET sample; and consideration of γγ attenuation, significant above ~ 10 GeV for redshifts z > 3–5. We find that GLAST should detect > 200 GRBs per year, with sensitivity to a few tens of GeV for a few bursts. GLAST could detect the energy- and distance-dependent dispersion (10 ms / GeV / Gpc) predicted by QG with 1–2 years of observations. Attribution to QG would require correlation of GRB redshifts with the temporal and energetic signatures.

## 1 Introduction

Since multi-wavelength strategies for observing gamma-ray bursts (GRBs) were coordinated in 1997, virtually all GRBs observed simultaneously with gamma-ray detectors and X-ray telescopes have revealed prompt X-ray transients and afterglows. X-ray counterparts, with few arc minute accuracies, have enabled deep optical searches, and for approximately half of the events an optical counterpart has been discovered. Presently, seven redshifts have been obtained, associated with absorption of the optical transient's light, or with the host galaxy after the transient has faded: z = 0.695, 0.835, 0.958, 0.967, 1.096, 1.600, and 3.412 (see e.g., Djorgovski 1999). For $\{\Omega_m, \Omega_\Lambda\} = \{0.3, 0.7\}$ and $H_0 = 65$ km s$^{-1}$ Mpc$^{-1}$, these redshifts represent co-moving distances $d_{CM}$ ~ 3–7 Gpc. Possible selection effects against seeing optical transients and therefore measuring a higher fraction of GRB redshifts include: lower intrinsic optical luminosities, more rapid fading, higher optical depth in the source's environment, and larger GRB source distances. However, with the present samples of redshifts and gamma-ray fluxes, we can infer that the median redshift is at least of order unity, and that the (would-be isotropic) luminosity distribution spans ~ $10^{52}$ – $3 \times 10^{54}$ erg s$^{-1}$. However, the optical decay in the most luminous case, GRB 990123 with z = 1.6, suggests that the gamma radiation is beamed into a narrow solid angle, $\Omega = \varepsilon\, 4\pi$, with $\varepsilon$ ~ 0.02 (Kulkarni et al. 1999). Still, GRB 990123 is the brightest GRB with a determined redshift, corresponding to a co-moving distance of 4.9 Gpc.

Here we describe how the supra-GeV radiation in narrow pulses of the brightest GRBs propagating over cosmological distances, and observed by the proposed Gamma-ray Large Area Space Telescope (GLAST), can be utilized to provide significant tests of energy-dependent dispersion, predicted in some realizations of quantum gravity (QG). In particular, Amelino-Camelia et al. (1998) suggest that quantum-gravitational fluctuations give rise to a velocity dependence that may be observable in GRB time profiles,

$$v \approx c\,[1 - \xi\, E_\gamma / E_{QG}] \quad (1)$$

where $E_\gamma$ is the photon energy, and $E_{QG}$ ~ $10^{19}$ GeV, believed to be associated with the Planck energy of unification. Whether the predicted effect would be tachyonic ($\xi = -1$), or not, is uncertain (J. Ellis, priv. comm.). For GeV energies, equation (1) translates into time lags of order

$$\Delta t \approx 10\text{ ms} \times [E_\gamma / 1\text{ GeV}]\, [d_{CM} / 1\text{ Gpc}]. \quad (2)$$

Since individual pulses in GRBs have well-measured widths of order 100 ms at energies of a few × 100 keV, narrowing at higher energies, and since pulse rises are faster than decays (Norris et al. 1996), the onsets of GRB pulses in the MeV regime should provide excellent fiducial markers for measuring the energy-dependent dispersion expected at GeV energies. Recently, Biller et al. (1998) set a limit on $E_{QG} > 4 \times 10^{16}$ GeV, from TeV observations of AGN variability on timescales of ~ $10^3$ s. Improved limits will probably require faster phenomena, like GRBs, which lie at distances comparable to those of AGN.

It remains to demonstrate that GLAST will detect sufficient numbers of photons at ~ 100 MeV to GeV energies to make such a project feasible. In the next section we briefly discuss the capabilities of GLAST relevant to the GRB-QG measurement, and then describe simulations of GRBs detected by GLAST. In the last section we describe the expected statistics in terms of the average pulse profile as a function of energy.

## 2 GRBs and GLAST

Figure 1 illustrates GRB 930131, the "Superbowl burst," an extreme example detected by *Compton* GRO's BATSE, and in the EGRET field of view. The event comprised an extremely fast and intense initial pulse (with significant substructure at BATSE energies: see Kouveliotou et al. 1994) accompanied by lower emission lasting for seconds. GRB 930131 belongs to the short variety of bursts, with $T_{90}$ durations (time to accumulate 5% to 95% of counts) less than 2 s. No optical transients, and therefore no redshifts, have been found for short bursts so far. However, their pulses tend to be a factor of 10–20 shorter than pulses in long bursts (total durations: tens to a few hundreds of seconds), and so short bursts would provide more precise temporal markers if their source distances could be reliably calibrated in some manner.

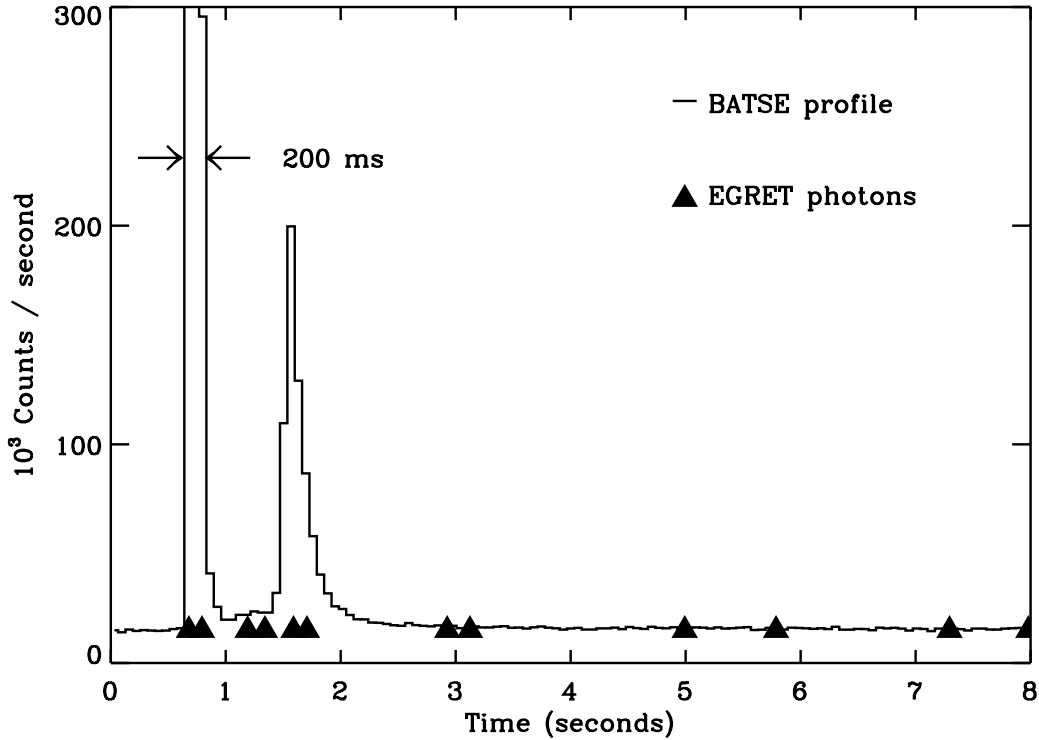

**Figure 1:** The intense, short burst of 930131, whose wavefront passed the Earth during the Superbowl. The dead-time corrected estimate of the maximum BATSE count rate (> 25 keV) is ~ $2 \times 10^6$ s$^{-1}$. Energies of the two EGRET photons (triangles) during the initial spike are ~ 30 and 80 GeV. EGRET dead time is ~ 100 ms per event, hence the actual high-energy rate during first pulse is very uncertain.

GRB fluxes expected from GLAST (Atwood et al. 1999) should be considerably higher than those seen by EGRET, for five reasons: (1) GLAST will have 5–6 times the effective area compared to EGRET; (2) the pattern recognition algorithms and instrument are being designed to defeat gamma self veto (arising from calorimeter backsplash, important at energies above 1 GeV) in the anti-coincidence system; (3) the calorimeter-only signal from GRBs may be usable, thus multiplying the 0.5-radiation length tracker signal by a factor of $\sim 2.5$; (4) GLAST will have negligible dead time. In Figure 1, the triangles represent times of arrival of photons detected in EGRET's spark chamber, which has $\sim 100$ ms dead time per event, comparable to pulse widths at BATSE energies. Thus, only two photons could be detected by EGRET during the initial spike of GRB 930131. And (5) GLAST will have a usable field of view (FOV) for GRBs of at least $\pi$ steradians, $\sim 5$ times that of EGRET. The effective area of GLAST is $\sim 3$ m$^2$, with a tracker depth of $\sim 60$ cm, and a calorimeter depth of $\sim 9$ radiation lengths of CsI.

We performed in-depth simulations of GRB observations by GLAST, using a C++ code which embodies a detailed geometric and material characterization of the instrument; electromagnetic (EGS4) and hadronic (Gheisha) interactions; and pattern reconstruction algorithms developed for GLAST. The GRB simulations were detailed previously in Bonnell et al. (1997a). The present results were obtained from a similar treatment, the main departure being adoption of a Gaussian distribution of spectral power-law indices, $\alpha = 2.0 \pm 0.2$, which appears to represent the current sample of average EGRET GRB spectra observed with its calorimeter to $\sim 200$ MeV (J. Catelli, priv. comm.). We recapitulate the most significant aspects here.

The fluence distribution of power-law spectrum GRBs above 100 MeV was assumed to be proportional to the observed BATSE peak-flux distribution (log $N$ - log $P$) and to a brightness-independent duration distribution (Bonnell et al. 1997b). The distribution of GRB spectra was normalized by scaling the BATSE peak-flux distribution to yield the flux (photons cm$^{-2}$ s$^{-1}$ MeV$^{-1}$) at 1 MeV. The BATSE log $N$ - log $P$ was constrained to follow a -3/2 power law at the bright end, avoiding biases due to small number statistics for detected bright bursts. The photon number spectrum was then integrated above 100 MeV to yield the number of photons per second for a power-law index chosen randomly from the distribution. The number of photons was scaled by a duration chosen from the brightness-independent distribution, to yield the fluence (gammas > 100 MeV cm$^{-2}$) incident normally on GLAST. Calorimeter-only detected photons were accepted above 1 GeV, and excluded at lower energies due to increased background within the photon localization radius. Bursts were randomly distributed on the sky (BATSE would detect 815 bursts yr$^{-1}$, full sky), and deemed detectable by GLAST within an acceptable angle of 75° with a zenith-pointed mode. GLAST's PSF and effective area were adjusted per random burst direction. Approximately 230 simulated bursts were detected by GLAST yr$^{-1}$, a result that is strongly dependent on average spectral index.

## 3 GRB Pulse Shape and Statistics with QG

We then constructed the "average" GRB pulse versus photon energy that GLAST should see above 100 MeV under the following four assumptions: (1) Only photons from long bursts ($T_{90} > 2$ s) can be utilized, and strictly, only $\sim 25\%$ of long bursts – those with optical counterparts *and* with determined redshifts – would contribute. (2) Pulses at GLAST energies (10 MeV – 100 GeV) have widths comparable to pulses observed at the limit of BATSE energies ($\sim 1$ MeV), $\sim 50$–100 ms. (3) Only one-fifth of the flux resides in identifiable, separable pulses found at MeV energies; this restriction may be too conservative, given that pulse recognition techniques such as described in Scargle (1999) are improving our ability to sort through GRB time profiles. (4) A GRB monitor accompanies GLAST, providing well-determined pulse shapes in the canonical GRB band, 100 keV – few MeV, which serve as fiducial temporal markers to align the GeV pulses in registration. This also presupposes a well-planned GRB alert system and dedicated optical follow-up program. With the above assumptions, Figure 2 illustrates the composite exponential pulse (FWHM constant: 50 ms) from 20 bright bursts accumulated over 2 yrs, with photons distributed randomly except for application of QG dispersion (setting $\xi = +1$ in equation [1]), and placing all 20 GRB sources at 1 Gpc.

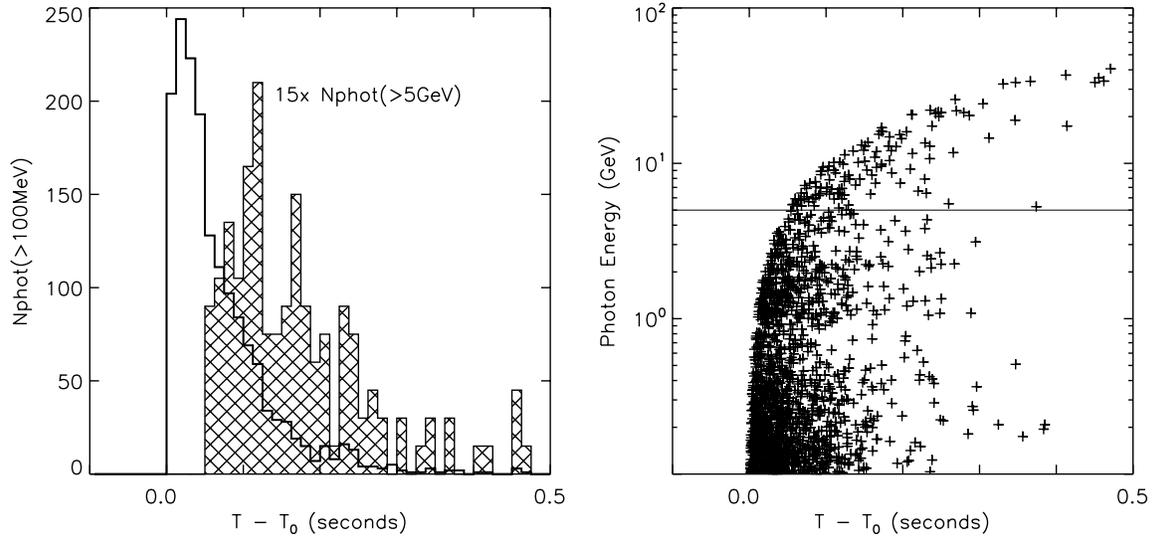

**Figure 2**: Left – Composite pulse includes GLAST-detected photons from 20 bright GRBs observed over 2 years. Energy threshold cuts are 0.1 GeV (open histogram) and 5 GeV (cross-hatched). Right – Arrival times versus photon energy. Horizontal line indicates 5 GeV threshold. Dispersion due to QG is evident.

Photon statistics at the higher energies will be diminished at high redshift by intergalactic γγ attenuation. A recent treatment by Stecker (1999) predicts optical depth $\tau \sim 1$ at 10 (20) GeV for $z \sim 5$ (2). Mannheim, Hartmann & Funk (1996) predict $\tau \sim 1$ at 10 GeV for $z \sim 3$. Determined GRB redshifts put most sources at $z < 3$; hence we expect the total usable photon counts up to $\sim 10$ GeV to be fairly well represented in Figure 2. Regardless of the nuances in these treatments, above 10 GeV, the intrinsic GRB flux decreases markedly. Thus the usable energy range for probing QG with GLAST is limited below, by pulse registration versus dispersion timescales, to > 10 MeV, and above by intrinsic photon statistics and extrinsic γγ attenuation. In conclusion, even after considering intergalactic γγ attenuation, GLAST should be able to see the QG dispersion effect described in equation (1). Correlation of photon lag with GRB distance is an additional expectation of the QG framework. The observed range in $d_{CM}$, 3–7 Gpc, with a few closer bright bursts expected to be observed over long observation intervals, should reveal the dependence of QG lag on distance. Note that another extrinsic factor, cosmic time dilation, stretches GRB pulses by (1+z).

# References

Amelino-Camelia, G., et al. 1998, Nature 393, 763
Biller, S.D., et al. 1998, Phys. Rev. Lett., submitted, gr-qc/9819944
Bonnell, J.T., et al. 1997a, Proc. 25[th] ICRC (Durban) 3, 69
Bonnell, J.T., et al. 1997b, ApJ 490, 79
Djorgovski, S.G. 1999, in ITP conference "GRBs and Their Afterglows", March 15-19, Santa Barbara
Kouveliotou, C., et al. 1994, ApJ 422, L59
Kulkarni, S., et al. 1999, Nature, submitted, astro-ph/9902272
Mannheim, K., Hartmann, D., & Funk, B. 1996, ApJ 467, 532
Norris, J.P., et al. 1996, ApJ 459, 393
Atwood, W.B., et al. 1999, Nucl. Instr. Meth., submitted
Scargle, J.D. 1998, ApJ 504, 405
Stecker, F.W. 1999, Proc. of DPF '99, in press, astro-ph/9904416